\documentclass[pra,twocolumn,showpacs,preprintnumbers,groupedaddress,amsmath,amssymb]{revtex4}
\usepackage{hyperref}
\usepackage{graphicx}
\usepackage{dcolumn}
\usepackage{bm}

\begin{document}
\title{See-saw protocol for Doppler cooling of multilevel systems with coherent pulse trains}
\author{Mahmoud Ahmad}
\affiliation{Department of Physics, University of Nevada, Reno, Nevada 89557, USA}
\author{Ekaterina Ilinova}
\affiliation{Department of Physics, University of Nevada, Reno, Nevada 89557, USA}
\author{Andrei Derevianko}
\affiliation{Department of Physics, University of Nevada, Reno, Nevada 89557, USA}

\begin{abstract}
We consider decelerating and cooling an ensemble of multilevel atoms and molecules with a coherent train of ultrashort laser pulses. In the frequency domain such trains form frequency combs.  We  propose a novel see-saw scheme of Doppler cooling of multilevel atoms and molecules. In this scheme the teeth of the frequency comb are periodically moved in and out of resonance with the allowed transitions. The see-saw cooling may be carried out in practice by switching carrier-envelope phase offset between predefined values. We study the performance of the see-saw scheme for the simplest prototype  three-level $\Lambda$ system and demonstrate its advantage. For these systems we demonstrate a procedure for determining  optimal train parameters, and see-saw switching period. We also illustrate the performance of the protocol by numerically simulating
time evolution of velocity distribution.
\end{abstract}

\pacs{37.10.De, 37.10.Gh, 42.50.Wk}
\maketitle

\section{Introduction}
\label{sec:Introduction}

The most widespread method of direct laser cooling of atoms is Doppler cooling \cite{HanSch75},
in which a sample of atoms is irradiated with laser light.  Each atom absorbs laser photons and then radiates the light in random directions  returning to the ground state. Repeating this optical cycle typically tens of thousands of times  can cool the atomic sample down to the mK temperatures.
Direct application of the conventional Doppler cooling to multilevel atoms and molecules is challenging: most atoms and all molecules have transitions that can radiatively branch out to a multitude of other states. Exciting population from all
these lower-energy states requires a large number of lasers which makes the conventional scheme impractical.

 Direct cooling of only a few special multilevel atomic and molecular systems have been demonstrated so far.  Recently the cooling of multilevel atoms Er and Dy were experimentally demonstrated \cite{YouLuLev10,MinBurLev12}.  Doppler cooling of a certain class of a molecules with highly diagonal Franck-Condon matrixes has been demonstrated \cite{ShuBarGle09}. The prospects for laser cooling of TlF molecules have been reported \cite{HunPecGre12}.

The enumerated multilevel systems have very unique properties; here we address the possibility of cooling general multilevel   system.
The basic idea is to employ a frequency comb (FC). Frequency combs are produced by coherent pulse train emitted by mode-locked lasers. Their frequency spectrum consists of a regular comb of sharp lines, i.e., effectively a large collection of CW lasers. By moving the teeth in resonance with individual transitions one can selectively
transfer populations between the levels, thereby enabling cooling  multilevel
systems.

We start by introducing basic parameters of pulse trains  and frequency combs. The electric field of the train may be expresed as

\begin{equation}
\mathbf{E}(t)=\hat{\varepsilon}\,E_{p}\,\sum\limits_{m}\cos(\omega_{c}%
t+\Phi_{m})\,g(t-mT)
\label{Eq:TrainField} \, ,
\end{equation}
where $\hat{\varepsilon}$ is the polarization vector, $E_{p}$ is the field amplitude,  $\Phi_{m}$ is the phase shift, $\omega_c$ is the carrier laser frequency and $T$ is the pulse repetition period.  $\Phi_m=m\times\varphi$ is the accumulated phase with $\varphi$ being the carrier envelope phase offset (CEO) between subsequent pulses. The pulse envelope $g(t)$ is normalized so that $\max g(t)\equiv1$. The spectral width of a single mode is $\sim\sqrt{24}/(NT)$, where $N$ is the number of pulses in the train \cite{T.Haensch2007}. For an infinite train, $N\rightarrow\infty$, the spectral profile of the pulse train exhibits a characteristic periodic structure composed of a series of regularly spaced laser modes
\begin{equation}\label{Eq:FC_modes}
\omega_n=\omega_c+\frac{2n\pi}{T} +\frac{\varphi}{T},
\end{equation}
where $n$ is the index of the mode.
The frequency interval between the neighbouring modes is determined by the pulse repetition rate $\omega_{rep}=2\pi/T$.  The third term, i.e., $\varphi/T$, shifts the entire FC while keeping intervals between teeth constant.

\begin{figure}[h]
\begin{center}
\includegraphics*[width=3.2in]{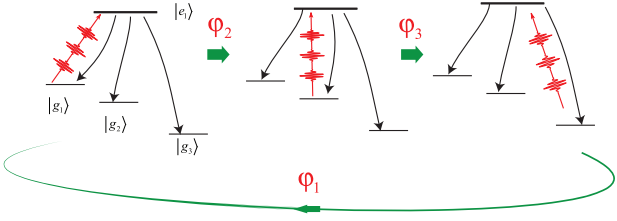}
\end{center}
\caption{ See-saw cooling protocol of multi-level system with a tunable frequency comb.}
 \label{Fig:protocol}
\end{figure}

Our method is based on varying the carrier-envelope phase offset $\varphi$ (see Eq.~(\ref{Eq:FC_modes})).
We alter the pulse train in such a way that the teeth move in and out of resonance with specific transitions. The time-sequence is shown in Fig.~\ref{Fig:protocol}. For illustrative purposes, we consider a four-level system with three ground states ($| \text{g}_1 \rangle$,$| \text{g}_2 \rangle$,$| \text{g}_3 \rangle$) and one
excited state $| e_1 \rangle$. Suppose the atom/molecule is initially in the $| \text{g}_1 \rangle$ state. We start by tuning the comb to
the frequency of the $| \text{g}_1 \rangle-| e_1 \rangle$ transition. We Doppler cool the atom on this transition. At each radiative emission event a
part of the population branches out to the $| \text{g}_2 \rangle$ and $| \text{g}_3 \rangle$ levels. When the population
of the $|\text{g}_1\rangle$ level is depleted, by varying the carrier-envelope phase $\varphi$ we tune the comb to the frequency of  the  $| \text{g}_2 \rangle-| e_1 \rangle$ transition and cool on that transition. Again due to the radiative branching, the population will leak to the $|\text{g}_1\rangle$ and $|\text{g}_3\rangle$ levels. We then retune the comb to the $| \text{g}_3 \rangle-| e_1 \rangle$ transition and cool on that transition. And so on.
As a result the atom or molecule is cooled with a single laser source. We refer to this method as the see-saw scheme.

\section{See-saw Doppler Cooling applied to the $\Lambda$-system}
\begin{figure}
\begin{center}
\includegraphics*[width=3.4in]{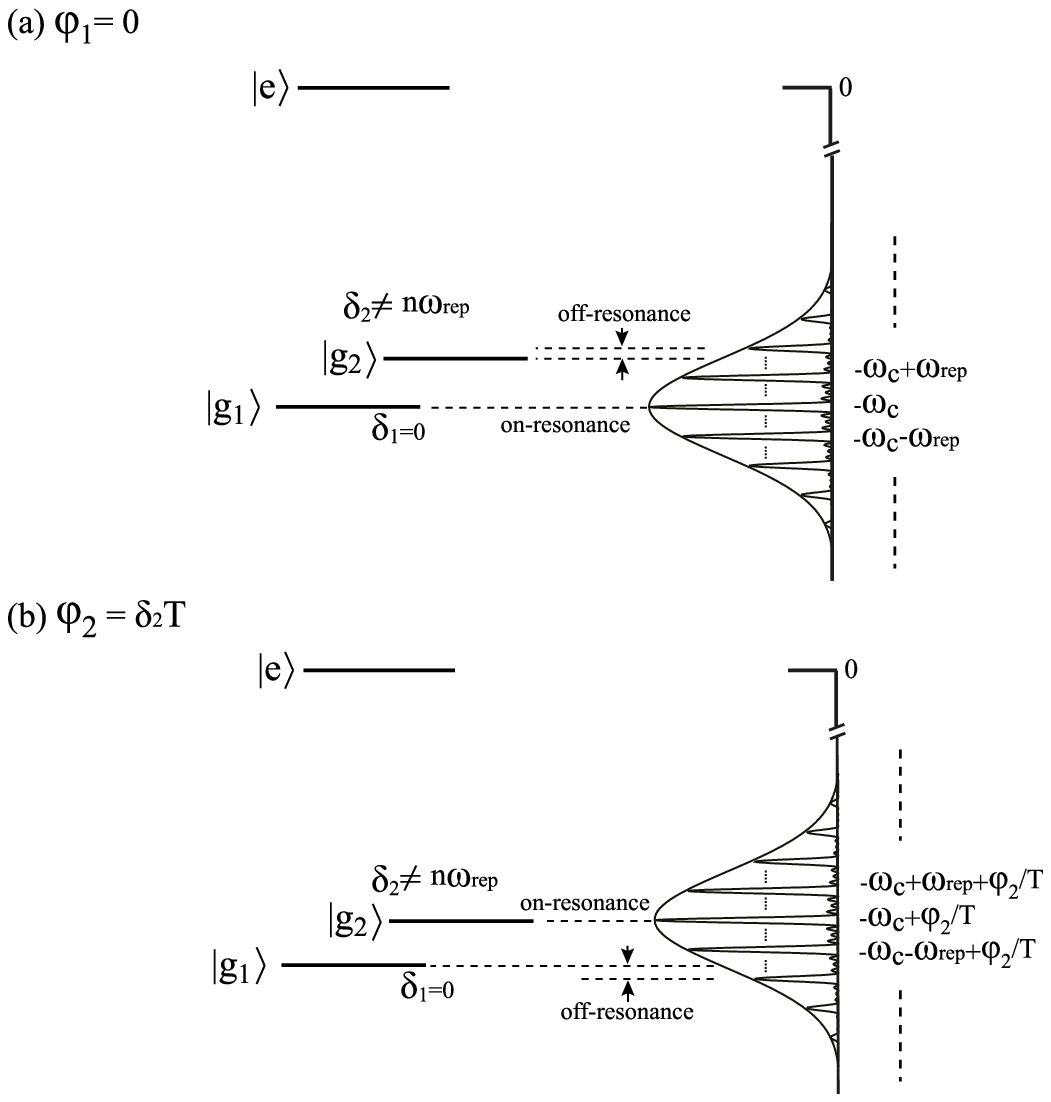}
\end{center}
\caption{See-saw cooling protocol applied to the $\Lambda$ system. (a) Shows the energy level frequencies relative to the
FC teeth positions for $\varphi=\varphi_1$. (b) Shows new detunings of the ground states when the CEO phase in the subsequent period is set to
$\varphi=\varphi_2$. Alternating the phase between $\varphi_1$ and $\varphi_2$ moves FC teeth in and out of resonance with either $| \text{g}_1$ or $| \text{g}_2 \rangle$.}
\label{Fig:Opt-scheme}
\end{figure}

\begin{figure}
\begin{center}
\includegraphics*[width=3.2in]{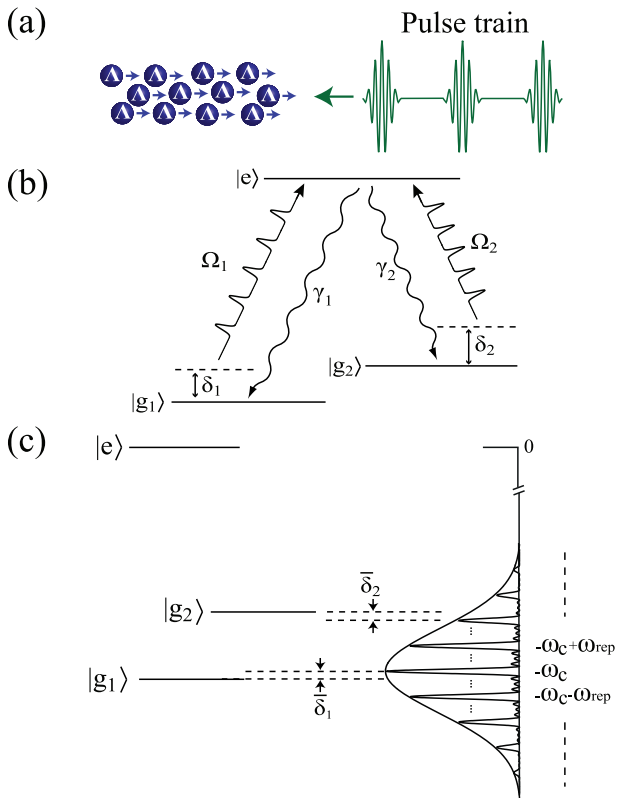}
\end{center}
\caption{(a) An atomic beam of $\Lambda$-type systems interacting with a counter propagating pulse train. (b) Energy
level diagram of $\Lambda$ system. (c) Frequency domain picture of comb interacting with $\Lambda$ system. }
\label{Fig:3LVL_scheme}
\end{figure}

Below we demonstrate how the proposed see-saw cooling scheme can be applied for cooling the simplest prototype multilevel system: three-level atoms. We limit our consideration to the $\Lambda$ configuration, as sketched in Fig.~\ref{Fig:3LVL_scheme}(b).
The system has two ground states
$|\text{g}_1 \rangle$ and $|\text{g}_2\rangle$, and one excited state $|\text{e} \rangle$.
The energy gap between the two ground states is $\Delta_{12}$. The excited state lifetime $\tau=1/\gamma$, where the total radiative decay rate is the sum of the decay rates to the two ground states $\gamma=\gamma_1+\gamma_2$.  The Rabi-frequencies of individual transitions are denoted as $\Omega_1$ and $\Omega_2$.

The atomic beam is irradiated by the counter-propagating train of
laser pulses Eq.(\ref{Eq:TrainField}) as illustrated in Fig.~\ref{Fig:3LVL_scheme}(a).

The optical Bloch equations for the density matrix governing the evolution of populations $\rho_{ee}$, $\rho_{g_jg_j}$ and coherences $\rho_{eg_j}$, $\rho_{g_1g_2}$ ($j=1,2$) read
\begin{eqnarray}\label{Eq:OBE1}
\dot{\rho}_{ee}&=&-\gamma\rho_{ee}-\sum\limits_{j=1}^2Im\left[\Omega_{eg_j} \rho_{eg_j}\right],\\
\dot{\rho}_{eg_j}&=&-\frac{\gamma}2\rho_{eg_j}+i \sum\limits_{j=1}^{2} \frac{\Omega^*_{eg_j}}2 (\rho_{ee}\delta_{jp}-\rho_{g_pg_j}), \label{Eq:OBE2}\\
\dot{\rho}_{g_jg_{j'}}&=&\delta_{jj'}\gamma_j \rho_{ee}+\frac{i}2(\Omega^*_{eg_{j'}}\rho_{g_je}-\Omega_{eg_j}\rho_{eg_{j'}}).\label{Eq:OBE3}
\end{eqnarray}
The time- and space-dependent Rabi frequency is
\begin{equation}\label{Eq:rabifreqsp}
\Omega_{eg_j}(z,t)=\Omega^{peak}_j \, \sum_{m=0}^{N-1} g(t+\frac{z}{c}-m T )e^{-i(k_cz(t)-\delta_jt-\Phi_m)},\,
\end{equation}
where $\delta_j=\omega_{c}-\omega_{eg_j}$, $k_c=\omega_c/c$ and $z$ is the
atomic coordinate. The peak Rabi frequency is $\Omega^{peak}_j=\frac{E_{p}}{\hbar}\langle e|\mathbf{D}\cdot\hat{\varepsilon}|g_j\rangle$. Eqs.~(\ref{Eq:OBE1}, \ref{Eq:OBE2}, \ref{Eq:OBE3}) were derived using the rotating wave approximation. The pulse areas corresponding to individual transitions are: $\theta_i=\Omega^{peak}_i\int_{-\infty}^\infty g(t)dt$.
The residual detunings of transitions $|g_i\rangle\rightarrow|e\rangle$ from the nearest FC teeth, $\overline{\delta}_i$ can be expressed as
\begin{equation}
\overline{\delta}_i=(\bar{\eta}_{i}+2\pi n_i)/T,
\end{equation}
where \begin{equation}\bar{\eta}_{i}=(\delta_i+k_cv)T+\varphi \label{Eq:eta}\end{equation}
is the Doppler-shifted phase offset between the subsequent pulses \cite{IliAhmDer11, IliDer12b}, $k_c$ is the wave vector and an integer $n_i$ is chosen in order to normalize the detunings to the $-\omega_{rep}/2\leq\bar{\delta}_i<\omega_{rep}/2$ interval.

Below we assume that individual pulse areas and radiative decay rates are equal:  $\theta_1=\theta_2$, $\gamma_1=\gamma_2$.
Direct population transfer between the two ground states is dipole forbidden. Therefore, the second
ground state is accessible only through the excited state. Additionally, the atoms are assumed to be initially prepared in the first
ground state $|\text{g}_1 \rangle$.

The scattering force exerted on atoms by the pulse train can be expressed in terms of the coherences as \cite{IliDer12b}
\begin{equation}\label{ClFrc}
F_z=\hbar k_c \sum\limits_{j=1}^2\mathrm{Im}[\rho_{eg_j}\Omega_{eg_j}] \, .
\end{equation}
In the limiting case when the duration of the pulse is much shorter than the repetition
period $T$, one can utilize the delta-function pulse shape model to describe the dynamics of the  system \cite{Ilinova2011}.
In this approximation, the fractional momentum kick due to the interaction of an atom with the $n$-th pulse reads
\begin{equation}\label{Eq:Dp/Ppr}
\Delta p_{n}/p_r=\left(\rho_{ee}^{n}\right)_r-\left(\rho_{ee}^{n}\right)_l,
\end{equation}
where $p_r=\hbar k_c$ is the recoil momentum, and $\left(\rho_{ee}^{n}\right)_{l,r}$ are the populations of the exited state right before
and after the $n$-th pulse.

The average scattering force may be determined as
\begin{equation}
F_{avg}=\frac{\Delta p}{N T},
\end{equation}
where $\Delta p$ is the total momentum imparted by the pulse train, $T$ is its repetition period and $N$ is the number of pulses in the train.
Maximum value of $F_{avg}=p_r/T$ is attained when each pulse transfers full recoil momentum to the atom.


We suppose that initially the central FC tooth is resonant with the $|e\rangle\rightarrow|\text{g}_1\rangle$ transition.
Then the frequency of the $|e\rangle\rightarrow|\text{g}_2\rangle$ transition will be detuned with respect to the central  tooth by $-\Delta_{12}$, where $\Delta_{12}$ is the energy gap between the two ground states.
The nearest to the $\omega_{eg_2}$ tooth will be detuned from it by $mod(\Delta_{12},\omega_{rep})$ if $mod(\Delta_{12},\omega_{rep})<\omega_{rep}/2$ and $mod(\Delta_{12},\omega_{rep})-\omega_{rep}$ if $mod(\Delta_{12},\omega_{rep})>\omega_{rep}/2$.  In the first case it is blue detuned and in the second case it is red detuned.
The pulse repetition frequency is chosen so that $mod(\Delta_{12},\omega_{rep})\neq0$, that is the two photon resonance condition is avoided.
In case when $mod(\Delta_{12},\omega_{rep})=0$ the system evolves into the ``dark'' superposition of the two ground states which is transparent to the pulse train \cite{Ara04} and therefore does not exert the scattering force. The population of the system is initially in the first ground state $|\text{g}_1 \rangle$.
If the atoms are irradiated by the pulse train for some time, the population will migrate from the first, strongly driven,  ground state $|\text{g}_1\rangle$ to the second, less coupled, ground state $|\text{g}_2\rangle$.  Then according to the see-saw protocol, the phase offset can be switched $\varphi=\varphi_1=k_cv_{mp}T$ to $\varphi=\varphi_2$, where $\varphi_2=k_cv_{mp}T-mod(\Delta_{12},\omega_{rep})T$ if $mod(\Delta_{12},\omega_{rep})<\omega_{rep}/2$ and  $\varphi_2=k_cv_{mp}T+\omega_{rep}-mod(\Delta_{12},\omega_{rep})T$ if $mod(\Delta_{12},\omega_{rep})>\omega_{rep}/2$ ( where $v_{mp}$ is  usually the center of velocity distribution), so that the second transition becomes resonant with one of the FC modes.

\begin{figure}
\begin{center}
\includegraphics*[width=2.9in]{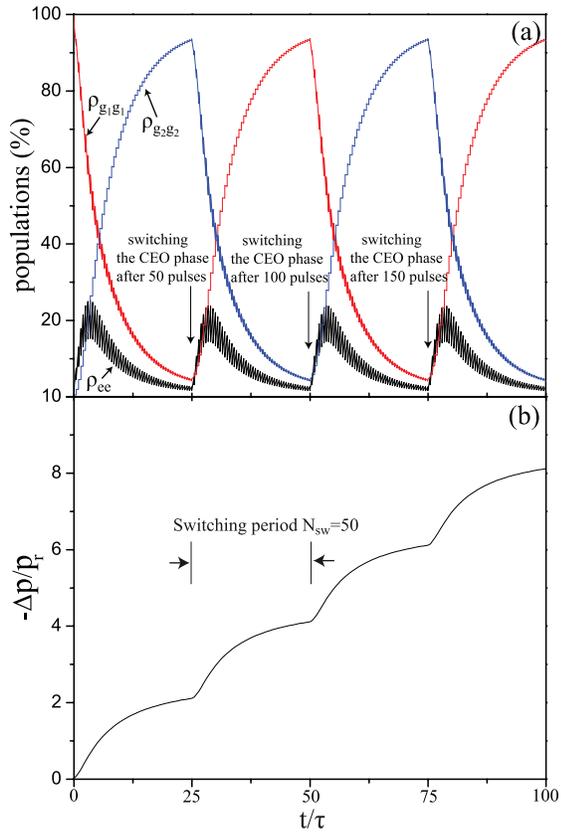}
\end{center}
\caption{(Color online) Time-dependence of $\Lambda$-system populations (panel (a)) and fractional momentum kick (panel (b)) in the see-saw protocol with $N_{sw}=50$.
These figures were computed by integrating OBEs, Eq. (\ref{Eq:OBE1}) with the following parameters: ground states splitting $\Delta=2\pi\times300\;\mathrm{GHz}$, spontaneous decays of $\gamma_{1}=\gamma_{2}=2\pi\times10$ $\mathrm{MHz}$, repetition period is $T=0.5/\gamma_e$ $\mathrm{ns}$, and pulse areas of $\theta_1=\theta_2=\pi/10$.}
\label{Fig:Populations_SW50PLS}
\end{figure}

To evaluate the time evolution of the density matrix we numerically integrated the optical Bloch equations (\ref{Eq:OBE1}, \ref{Eq:OBE2}, \ref{Eq:OBE3}). Fig.~\ref{Fig:Populations_SW50PLS}(a) shows the time evolution of population distribution over the first
100 lifetimes.
One can see that as the system approaches the quasi-steady state regime, the average population becomes redistributed mostly between the two ground states and the post pulse excited state population
saturates without further increase. To interrupt the saturation a change in the parameters is needed. This is
accomplished by changing the phase from $\varphi=\varphi_1$ to $\varphi=\varphi_2$. Such switching is illustrated in Fig.~\ref{Fig:Opt-scheme}(b).

\begin{figure}
\begin{center}
\includegraphics*[width=3.3in]{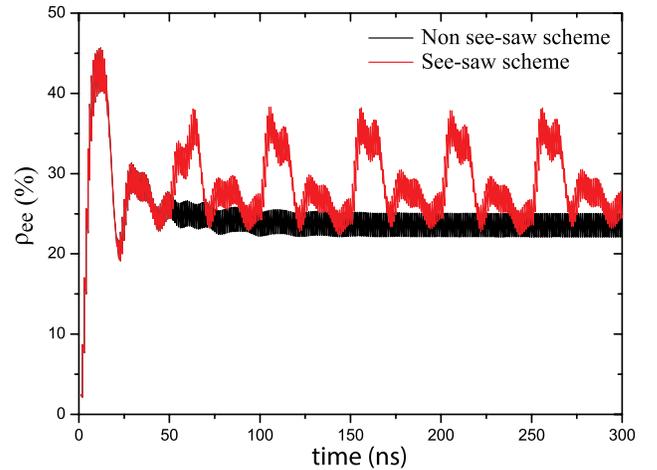}
\end{center}
\caption{(Color online)  Excited state  population as a function of time.
The black curve is for a fixed $\varphi$. The red curve is for see-saw scheme with $N_{sw}=50$.
The jump in the population is apparent once the phase is switched from $\varphi=0$ to $\varphi=-mod(\Delta_{12},\omega_{rep})T$.
These plots are  obtained for the following parameters: ground states splitting $\Delta_{12}=2\pi\times300$ $\mathrm{GHz}$, spontaneous decays of $\gamma_{1}=\gamma_{2}=2\pi\times10\;\mathrm{MHz}$,
repetition period is $T=1.00017\;\mathrm{ns}$, and pulse areas $\theta_1=\theta_2=\pi/10$.}
\label{Fig:No_seesaw_vs_seesaw}
\end{figure}

Fig.~\ref{Fig:Populations_SW50PLS}(b) shows the mechanical momentum accumulated by the atoms. It is clear that
at the end of each $50$-pulse period, the momentum accumulation saturates to a constant value. This means that the rate of
deceleration slows down. Once the phase is set to $\varphi=\varphi_2$ after the $50^{th}$ pulse,
a sudden increase in the accumulated momentum occurs which improves the cooling efficiency. Continuous
switching of the phase between $\varphi=\varphi_1$ and $\varphi=\varphi_2$ every $50^{th}$ pulse sustains more efficient
momentum transfer from the pulse train to the atomic beam.  Fig.~\ref{Fig:No_seesaw_vs_seesaw} compares the excited state dynamics for the non see-saw
and the see-saw schemes, switching the phase between $\varphi_1$ to $\varphi_2$ or vice versa every $50^{th}$ pulse (i.e., red curve). The average excited state population is increased compared to the case when the phase is fixed (i.e., black curve).

Hereafter in this paper, we denote $N_{sw}$ as the length of sub-train  for which we keep the phase constant at either values $\varphi_1$ and $\varphi_2$. For example $N_{sw}=50$ in Fig.~\ref{Fig:Populations_SW50PLS}.

\section{Cooling a 1D beam of $\Lambda$ systems}
\label{sec:SearchAlgorithm}
In this section we apply the proposed see-saw cooling method for cooling a thermal 1D beam of three-level $\Lambda$-type
atoms.  First we find the optimal parameters maximizing the scattering force exerted on the atoms.
Second we study the evolution of the ensemble velocity distribution during cooling with these parameters.

\subsection{Maximum scattering force}
The atoms under consideration have the $\Lambda$-like configuration of Fig.~\ref{Fig:3LVL_scheme}. Specific atomic values are taken to be
as follows: ground state splitting is $\Delta=2\pi \times 300\;\mathrm{GHz}$. Spontaneous decay rate to the ground sates are equal, i.e.,
$\gamma_{1}=\gamma_{2}=2\pi \times 10\;\mathrm{MHz}$, leading to the excited state lifetime of $8\;\mathrm{ns}$. Pulse areas are
taken to be equal as well, $\theta_1=\theta_2$. We optimize the FC parameters for the two different values of the pulse area
$\theta_{1,2}=\pi/10$ and $\pi/2$.

All parameters may be categorized into three groups
\begin{itemize}
  \item Atomic parameters: $\Delta_{12}$, $\gamma_{1}$ and $\gamma_{2}$.
  \item Fixed laser field parameters: $T$ and laser intensity.
  \item Variable laser-field parameter: $\varphi_{1}$, $\varphi_{2}$ and $N_{sw}$.
\end{itemize}
The description of the optimization process is summarized in the following text. Parameters $\Delta_{12}$, $\gamma_{1}$ and $\gamma_{2}$
remain fixed for the specific $\Lambda$ system. A mode-locked laser source is assumed to have fixed repetition period $T$ and intensity.
The intensity of the laser beam along with the transition dipole moments determine
the Rabi frequencies $\Omega_{1,2}$ and pulse areas $\theta_{1,2}$.
An optical element may be placed at the output of the laser source to control the CEO phases
$\varphi_{1}$ and $\varphi_{2}$, and the see-saw switching period $N_{sw}$.

We carry out an exhaustive search for the optimal values of $T$ and $N_{sw}$  corresponding to the maximum
possible average scattering force.
The maximum and minimum pulse repetition period values are taken as $T_{min}=1$ ns and $T_{max}=10$ ns. Maximum  and minimum  number of pulses between the switching are $N_{sw}^{min}=1$ and $N_{sw}^{max}=60$.
As an example, we obtained the optimal repetition periods and the optimal see-saw switching periods for the two pulse areas
$\theta_{1,2}=\pi/10$ and $\theta_{1,2}=\pi/2$.
For  $\theta_{1,2}=\pi/10$, optimal pulse repetition period was found to be $T^{opt}=1.0035\;\mathrm{ns}$ and $N_{sw}^{opt}=1$. For $\theta_{1,2}=\pi/2$,  $T^{opt}=1.0083\;\mathrm{ns}$ and $N_{sw}^{opt}=1$.

\subsection{Evolution of the velocity distribution}

\begin{figure}
\begin{center}
\includegraphics*[width=3.3in]{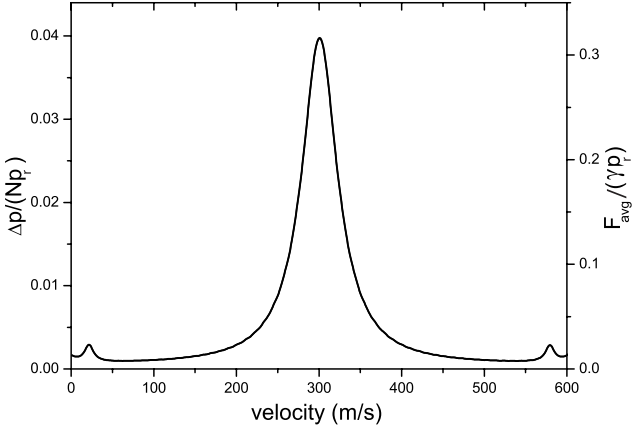}
\end{center}
\caption{Right axis is the average scattering force $F_{avg}$ plotted for the Doppler shifted velocities $v$ in the atomic beam. The left axis is
the average momentum kick per pulse $\Delta p/(Np_r)$ imparted to the atoms in the same atomic beam.}
\label{Fig:evoultionofvelocityfortheta=pi/10}
\end{figure}

\begin{figure}
\begin{center}
\includegraphics*[width=3.3in]{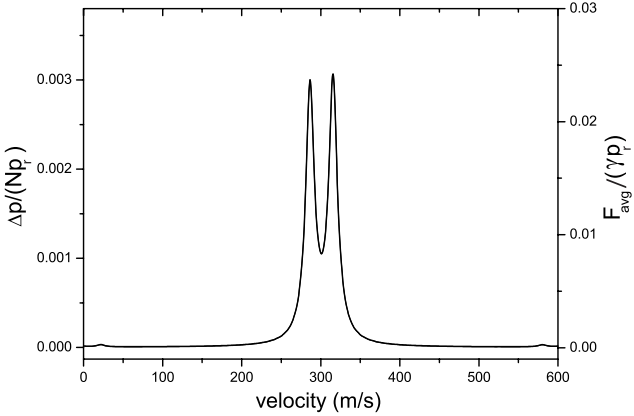}
\end{center}
\caption{The plot shows the two velocity dependent peaks of fractional momentum kick that were not resolved in Fig.~\ref{Fig:evoultionofvelocityfortheta=pi/10} due to
overlapping  of frequency comb teeth. The plot is for pulse areas $\theta_{1,2}=\pi/100$ otherwise the parameters are the same as Fig.~\ref{Fig:evoultionofvelocityfortheta=pi/10}.}
\label{Fig:evoultionofvelocityfortheta=pi/100}
\end{figure}

We consider a 1D atomic beam with the initial velocity distribution
\begin{equation}\label{Eq:initialVelDist}
    f(v,t=0)=\frac{9}{2}\frac{v^3}{v^4_{mp}}\exp{(-\frac{3v^2}{2v^2_{mp}})},
\end{equation}
where most probable value of the velocity $v_{mp}$ is set to $300\;\mathrm{m/s}$.

The carrier laser frequency is resonant with the atomic transition frequency, $\omega_c=\omega_{eg_1}$.
The values of the phase offsets $\varphi_1$, $\varphi_2$ are chosen with account of the Doppler shift of the laser frequency in the reference frame moving with the center of velocity distribution,  $\varphi_1=k_cv_{mp}T$, $\varphi_2=k_cv_{mp}T-mod(\Delta_{12},\omega_{rep})T$ (if $mod(\Delta_{12},\omega_{rep})<\omega_{rep}/2$), $\varphi_2=k_cv_{mp}T+\omega_{rep}T-mod(\Delta_{12},\omega_{rep})T$ (if $mod(\Delta_{12},\omega_{rep})>\omega_{rep}/2$). Initially  $\varphi_1=k_cv_{mp}T$ and as the atoms slow down $\varphi_1$ follows the $v_{mp}(t)$.

Fig.~\ref{Fig:evoultionofvelocityfortheta=pi/10} shows the fractional momentum
kick per pulse $\Delta p/(Np_r)$ versus atomic velocity for the pulse area $\theta_{1,2}=\pi/10$,  $T=1.0035\;\mathrm{ns}$
and $N_{sw}=1$. These optimal parameters correspond to the teeth separation of $\omega_{rep}\sim 2\pi \times 1\;\mathrm{GHz}$ in the frequency space and $v=\lambda_c/T\sim 587\;\mathrm{m/s}$ in the velocity space.

The FC teeth are initially resonant with the atomic transition frequency $\omega_{eg_1}$ for the atoms with velocities $v_{Dn}=v_{mp}+n\lambda_c/T$. Continuously as the atoms slow down the phases $\varphi_1$, $\varphi_2$
have to be shifted by the amount $-k_c\Delta v_{mp} T$ (where $\Delta v_{mp}$ is the decrease in the value of the central velocity due to the slowing of the beam) to compensate for the change in phases $\bar{\eta}_j$, Eq.(\ref{Eq:eta}). This leads to the formation of pronounced peaks in the velocity dependence of the scattering force,  separated by $\lambda_c/T\sim\;587\;m/s$. Fig.~\ref{Fig:evoultionofvelocityfortheta=pi/10} shows one
of such peaks for $n=0$ or $v_{D0}=v_{mp}=300\;\mathrm{m/s}$.

The velocity dependence of the scattering force can be explained on the basis of the Fourier transform of the see-saw pulse train Eq.\ref{Eq:TrainField}.
The Fourier transform (the derivation is given in the appendix) of the electric field (\ref{Eq:TrainField}) reads

\begin{widetext}
\begin{equation}
    \mathcal{E}(\omega)=F_1^+(\omega)\frac{\sin{(\frac{1}{2}N_{sw}\zeta_1)}}{\sin{(\frac{1}{2}\zeta_1)}}\times \sum_n\delta(\zeta-\frac{2n\pi}{N_{sw}})+F_2^+(\omega)\frac{\sin{(\frac{1}{2}N_{sw}\zeta_2)}}{\sin{(\frac{1}{2}\zeta_2)}}\times \sum_n\delta(\zeta-\frac{2n\pi}{N_{sw}}),
\end{equation}
\end{widetext}
where  $\zeta_i=\phi_i+(\omega-\omega_c)T$,
$\zeta=(\phi_1+\phi_2)+2(\omega-\omega_c)T$, and $F_{1,2}^\dagger(\omega)$ are the Fourier transforms of the pulse envelopes.

The arguments of the two $\delta$-functions determine positions of the FC teeth:
$\zeta-\frac{2n\pi}{N_{sw}}=0$ or $\omega_{n}=\omega_c+\frac{2n\pi}{2TN_{sw}}-\frac{\phi_1+\phi_2}{2T}$.
In the atomic reference frame  these values are Doppler shifted:
$\omega_{n}^a=k_cv+\omega_c+\frac{2n\pi}{2TN_{sw}}-\frac{\phi_1+\phi_2}{2T}$.
When  $\omega_{n}^a=\omega_{eg_i}$, then the $n$-th  tooth is resonant with the frequency of corresponding atomic transition for a given atomic velocity.

\begin{figure}
\begin{center}
\includegraphics*[width=3.3in]{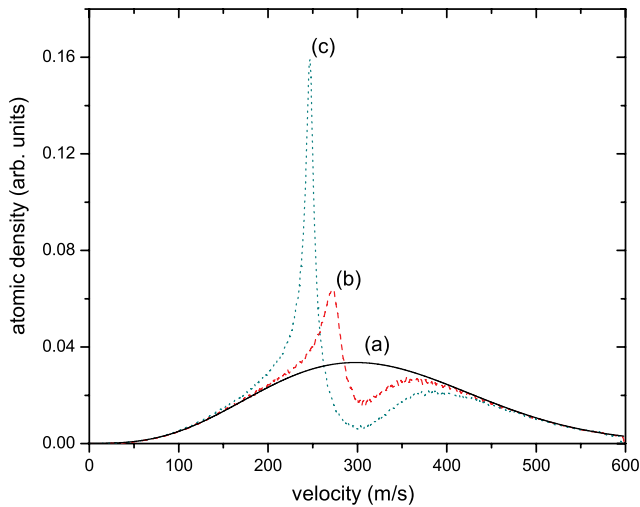}
\end{center}
\caption{Evolution of the velocity distribution of a 1D beam of $\Lambda$-type atoms interacting with a train of ultrashort laser pulses. Curve (a) displays the initial
velocity distribution of the atomic beam, curve (b) shows velocity distribution after interaction with 25,000 pulses, curve (c) shows velocity distribution evolution after
deceleration by 100,000 pulses. These were plotted for $\theta_{1,2}=\pi/10$, $T=1.0035\;\mathrm{ns}$ and $N_{sw}=1$. The remaining parameters are the same as in Fig.~\ref{Fig:evoultionofvelocityfortheta=pi/10}.}
\label{Fig:Atomic_Density_vs_velocity}
\end{figure}

Now we return to the interpretation of the velocity peaks in Fig.~\ref{Fig:evoultionofvelocityfortheta=pi/100}.
The carrier laser frequency is resonant with the frequency of the first atomic transition $\omega_c=\omega_{eg_1}$.
Then for a set of velocities $v_n^a=\left(\frac{\phi_1+\phi_2}{2T}-\frac{n\pi}{TN_{sw}}\right)/k_c$ the FC teeth will be resonant with the frequency of atomic transition $\omega_{eg_1}$.  At the same time for the velocities
 $v_n^b=\left(\frac{\phi_1+\phi_2}{2T}-\frac{n\pi}{TN_{sw}}-\Delta_{12}\right)/k_c$ the FC teeeth will be resonant with frequency $\omega_{eg_2}$ of another transition. The difference between the velocities $v_{n_1}^a$ and $v_{n_2}^b$ is determined by

 \begin{equation}
 v_{n_1}^a-v_{n_2}^b=\left(\frac{\pi}{TN_{sw}}(n_2-n_1)+\Delta_{12}\right)/k_c.
 \end{equation}
For $N_{sw}=1$ the minimum distance between the two resonant velocities equals to $mod(\Delta_{12},\omega_{rep})/k_c$.
If the pulse area is small ($\theta_i\ll1$) and the lifetime of the excited state is small compared to the pulse repetition period, then the two peaks corresponding to the nearest velocities $v_{n_1}^a$ and $v_{n_2}^b$ can be resolved, Fig.~\ref{Fig:evoultionofvelocityfortheta=pi/100}.

Figure~\ref{Fig:Atomic_Density_vs_velocity} shows time evolution of the atomic velocity distribution for various number of pulses. This is a numerical simulation
for a 1D atomic beam with the initial velocity distribution given by Eq.(\ref{Eq:initialVelDist}). We present two computed snapshots of velocity distribution for 25,000 and 100,000 pulses.
For the interaction with 25,000 pulses, there is a deceleration for a major part of the distribution accompanied with velocity compression. As the number of pulses interacting with atoms increases,  the continuous deceleration and the decrease of the width of velocity distribution is becomes more pronounced, as shown in Fig.~\ref{Fig:Atomic_Density_vs_velocity} curve (c).

\begin{figure}
\begin{center}
\includegraphics*[width=3.3in]{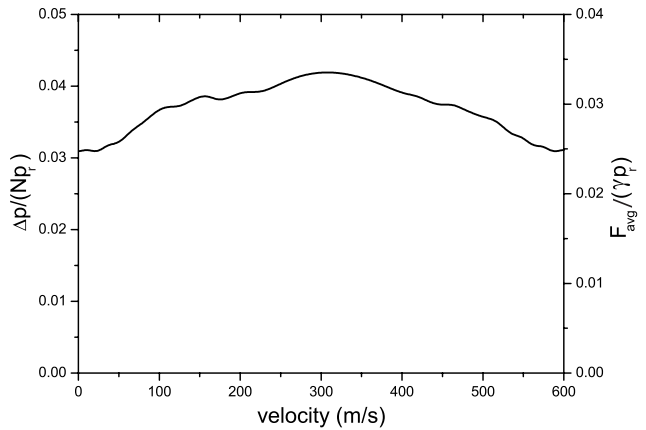}
\end{center}
\caption{Fractional momentum kick as a function of the Doppler shifted atomic velocity for repetition period $T=1.00083\;\mathrm{ns}$, switching period of $N_{sw}=1$,
and pulse areas $\theta_{1,2}=\pi/2$. Otherwise the parameters are the same as in Fig.~\ref{Fig:evoultionofvelocityfortheta=pi/10}.}
\label{Fig:dp_vs_v_theta1_eq_theta2_eq_pi_over_2_and_4}
\end{figure}

As the pulse area is increased,  the scattering force ceases to become velocity selective.
This is illustrated in Fig.~\ref{Fig:dp_vs_v_theta1_eq_theta2_eq_pi_over_2_and_4}  where we present the average momentum per pulse transferred to the atomic ensemble for single pulse areas $\theta_{1,2}=\pi/2$.  The optimal parameters for which the scattering force is maximized are:  $T=1.00083\;\mathrm{ns}$ and $N_{sw}=1$.

For the large pulse areas $\theta_{1,2}\sim\pi/\sqrt{2}$ the teeth in the scattering force spectral profile are essentially power-broaden. Low contrast of the maxima in velocity dependence of the scattering force leads to the slower compression of the velocity distribution.
As a result the ensemble can be decelerated, but not cooled.

\section{Conclusion}
We have proposed the see-saw protocol for decelerating and cooling multilevel systems by trains of ultrashort laser pulses.
We have demonstrated the efficiency of the proposed method in cooling the ensemble of three-level $\Lambda$-type atoms.
The method is based on periodic  interruption of the saturation regime in the system by switching the phase offset between subsequent pulses.  As a result the sustained  population transfer to the excited state occurs and nonzero scattering force is exerted on the atoms.
The see-saw scattering force is velocity-dependent and mimics the Fourier transform image of the see-saw pulse train.
At large single pulse areas $\Theta=\sqrt{\theta_1^2+\theta_2^2}\sim\pi$  the scattering force becomes largely velocity insensitive.
The same effect happens when the pulse repetition period is much longer  then the excited state lifetime.
The FC parameters, such as intensity, pulse repetition period and the number of pulses between the phase switchings can be optimized in order to maximize the scattering force. Additionally the phase offsets has to be adjusted during the deceleration process, while the position of the center of velocity distribution moves toward the lower velocities.

We illustrated the see-saw cooling by computing time evolution of velocity distribution for a thermal beam of three-level $\Lambda$-type atoms. The center of velocity distribution moves toward lower velocities, i.e., the beam is slowed down. Moreover the width of the distribution is decreased,  velocity distribution  is compressed, i.e., the ensemble in addition to being slowed is cooled.

\section{Acknowledgment}
This work was supported in part by the ARO and NSF.
\appendix*
\section{Fourier Transform of a Pulse Train with see-saw CEO phase switching}
\begin{figure}
\begin{center}\includegraphics[width=4in,height=2in]{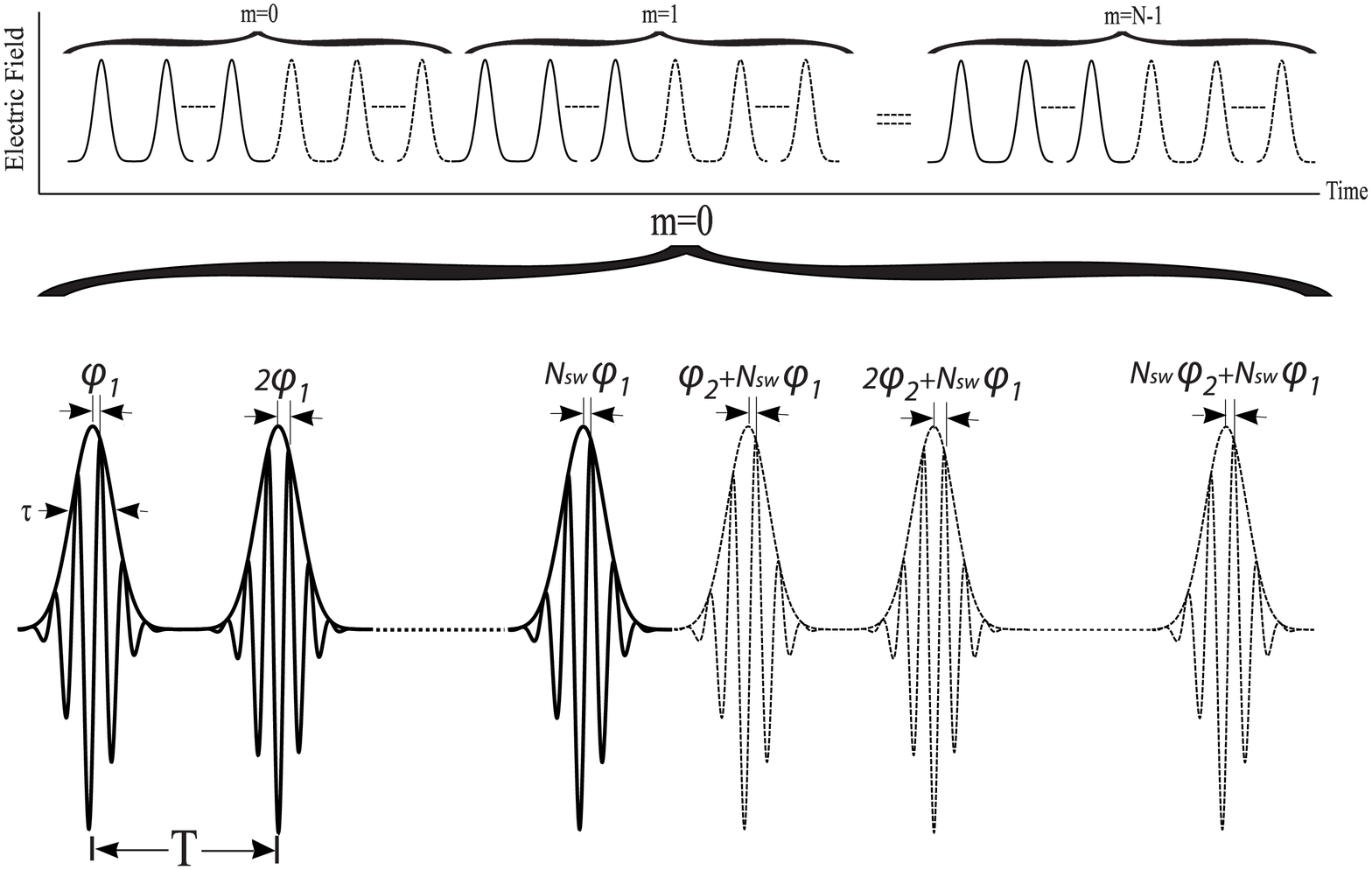}
\end{center}
\caption[Pulse train accumulating phases $\varphi_1$ and $\varphi_2$ alternatively every sequence of $N_{sw}$ pulses.]{The static picture of a pulse
train that accumulates phases $\varphi_1$ and $\varphi_2$ alternatively each $N_{sw}$ pulses.}
\label{ApndxFig:Efieldtrain}
\end{figure}

Consider a train of identical pulses emitted from an ideal mode-locked laser at equal time intervals. The repetition period between the pulses is fixed.
We consider the case when the pulses accumulates a CEO phase of $\varphi_1$ and $\varphi_2$ every $N_{sw}$ pulses alternatively. Fig.~\ref{ApndxFig:Efieldtrain}
illustrates such pulse train in the time domain. The snapshot of the pulse train in Fig.~\ref{ApndxFig:Efieldtrain} shows that the train may be
subdivided into $N$ shorter trains. Each shorter train has $2N_{sw}$ pulses. The first $N_{sw}$ pulses accumulate a CEO phase of $\varphi_1$ every pulse. The subsequent
$N_{sw}$ pulses accumulate a CEO phase of $\varphi_2$ every pulse. Therefore, the phase of a given pulse equals to the total accumulated phases for all the previous
individual pulses. The electric field at a fixed position for a pulse train of Fig.~\ref{ApndxFig:Efieldtrain} is
\begin{widetext}
\begin{align}
    \mathcal{E}(t)=&\mathcal{E}_1(t)+\mathcal{E}_2(t),\notag\\
    \mathcal{E}_1(t)=&\frac{1}{2}\mathcal{E}_0\sum_{m=0}^{N-1}\sum_{n=0}^{N_{sw}-1}(g(t)e^{-i\omega_ct+i(n+mN_{sw})\varphi_1+i(mN_{sw})\varphi_2}+ c.c.),\label{EtpT01}\\
    \mathcal{E}_2(t)=&\frac{1}{2}\mathcal{E}_0\sum_{m=0}^{N-1}\sum_{n=0}^{N_{sw}-1}(g(t)e^{-i\omega_ct+i(n+mN_{sw}+1)\varphi_2+i(mN_{sw}+N_{sw}-1)\varphi_1}+ c.c.).\label{EtpT02}
\end{align}
\end{widetext}
where $g(t)$ is the envelope function. The two terms in the preceding equation differ by the time shift of $N_{sw}T$ and their accumulative phases.
We compute the Fourier transform of Eq.(\ref{EtpT01}) and Eq.(\ref{EtpT02}) as $ \mathcal{E}(\omega)=\mathcal{F}[\mathcal{E}]=\frac{1}{\sqrt{2\pi}}\int_{-\infty}^{\infty}\mathcal{E}(t)e^{i\omega t}dt$.
The result with $\eta\equiv \omega-\omega_c$ reads
\begin{widetext}
\begin{align}
    \mathcal{E}(\omega)=&[F_1^+(\omega)\sum_{m=0}^{N-1}\sum_{n=0}^{N_{sw}-1}e^{i((n+2mN_{sw})\eta T+(n+mN_{sw})\varphi_1+mN_{sw}\varphi_2)}\notag\\
    &+F_2^+(\omega)\sum_{m=0}^{N-1}\sum_{n=0}^{N_{sw}-1}e^{i((n+2mN_{sw})\eta T+(n+mN_{sw})\varphi_2+mN_{sw}\varphi_1)}].\label{FT_of_the_train}
\end{align}
\end{widetext}
Here $F_1^+(\omega)$ is the Fourier transform of the first pulse of the first term in Eq.(\ref{EtpT01}) and $F_2^+(\omega)$ is is the Fourier transform of the first pulse
of the first term in Eq.(\ref{EtpT02}). For example, in case of a Gaussian envelope, $g(t)=e^{-t^2/2\tau_p^2}$,
\begin{align}
    F_1^+(\omega)=&\frac{1}{2}\tau\mathcal{E}_0 e^{-\frac{1}{2} \eta^2 \tau_p^2},\notag\\*
    F_2^+(\omega)=&\frac{1}{2}\tau\mathcal{E}_0 e^{-\frac{1}{2} \eta^2 \tau_p^2+iN_{sw}\eta T+i(N_{sw}-1)\varphi_1+i\varphi_2 }.\notag
\end{align}
In Eq.(\ref{FT_of_the_train}) we ignored the Fourier transform of the c.c. terms in Eq.(\ref{EtpT01}) and
Eq.(\ref{EtpT02}) as we are interested only in the frequency comb that is going to be built around $\omega_c$. By treating each sum in Eq.(\ref{FT_of_the_train}) as a geometric series,
we obtain

\begin{widetext}
\begin{equation}
    \mathcal{E}(\omega)=F_1^+(\omega)\frac{\sin{(\frac{1}{2}N_{sw}\zeta_1T)}}{\sin{(\frac{1}{2}\zeta_1T)}}\cdot\frac{\sin{(\frac{1}{2}N\;N_{sw}\zeta T)}}{\sin{(\frac{1}{2}N_{sw}\zeta T)}}+
    F_2^+(\omega)\frac{\sin{(\frac{1}{2}N_{sw}\zeta_2T)}}{\sin{(\frac{1}{2}\zeta_2T)}}\cdot\frac{\sin{(\frac{1}{2}N\;N_{sw}\zeta T)}}{\sin{(\frac{1}{2}N_{sw}\zeta T)}},\label{FTEtpT01}
\end{equation}
\end{widetext}
where $\zeta_i=\phi_i/T+(\omega-\omega_c)$ and $\zeta=(\phi_1+\phi_2)/T+2(\omega-\omega_c)$.
In the limit when $N\rightarrow\infty$, Eq.(\ref{FTEtpT01}) simplifies to
\begin{widetext}
\begin{align}
    \mathcal{E}(\omega)=F_1^+(\omega)\frac{\sin{(\frac{1}{2}N_{sw}\zeta_1T)}}{\sin{(\frac{1}{2}\zeta_1T)}}\cdot \sum_n\delta(\zeta-\frac{2n\pi}{N_{sw}T})+
    F_2^+(\omega)\frac{\sin{(\frac{1}{2}N_{sw}\zeta_2T)}}{\sin{(\frac{1}{2}\zeta_2T)}}\cdot \sum_n\delta(\zeta-\frac{2n\pi}{N_{sw}T}).
\end{align}
\end{widetext}


\begin{thebibliography}{10}
\expandafter\ifx\csname natexlab\endcsname\relax\def\natexlab#1{#1}\fi
\expandafter\ifx\csname bibnamefont\endcsname\relax
  \def\bibnamefont#1{#1}\fi
\expandafter\ifx\csname bibfnamefont\endcsname\relax
  \def\bibfnamefont#1{#1}\fi
\expandafter\ifx\csname citenamefont\endcsname\relax
  \def\citenamefont#1{#1}\fi
\expandafter\ifx\csname url\endcsname\relax
  \def\url#1{\texttt{#1}}\fi
\expandafter\ifx\csname urlprefix\endcsname\relax\def\urlprefix{URL }\fi
\providecommand{\bibinfo}[2]{#2}
\providecommand{\eprint}[2][]{\url{#2}}

\bibitem[{\citenamefont{Hansch and Schawlow}(1975)}]{HanSch75}
\bibinfo{author}{\bibfnamefont{T.}~\bibnamefont{Hansch}} \bibnamefont{and}
  \bibinfo{author}{\bibfnamefont{A.}~\bibnamefont{Schawlow}},
  \bibinfo{journal}{Opt. Comm.} \textbf{\bibinfo{volume}{13}},
  \bibinfo{pages}{68} (\bibinfo{year}{1975}).

\bibitem[{\citenamefont{Youn et~al.}(2010)\citenamefont{Youn, Lu, and
  Lev}}]{YouLuLev10}
\bibinfo{author}{\bibfnamefont{S.~H.} \bibnamefont{Youn}},
  \bibinfo{author}{\bibfnamefont{M.}~\bibnamefont{Lu}}, \bibnamefont{and}
  \bibinfo{author}{\bibfnamefont{B.~L.} \bibnamefont{Lev}},
  \bibinfo{journal}{Phys. Rev. A} \textbf{\bibinfo{volume}{82}},
  \bibinfo{pages}{043403} (\bibinfo{year}{2010}).

\bibitem[{\citenamefont{Lu et~al.}(2012)\citenamefont{Lu, Burdick, and
  Lev}}]{MinBurLev12}
\bibinfo{author}{\bibfnamefont{M.}~\bibnamefont{Lu}},
  \bibinfo{author}{\bibfnamefont{N.~Q.} \bibnamefont{Burdick}},
  \bibnamefont{and} \bibinfo{author}{\bibfnamefont{B.~L.} \bibnamefont{Lev}},
  \bibinfo{journal}{Phys. Rev. Lett.} \textbf{\bibinfo{volume}{108}},
  \bibinfo{pages}{215301} (\bibinfo{year}{2012}).

\bibitem[{\citenamefont{Shuman et~al.}(2009)\citenamefont{Shuman, Barry, Glenn,
  and DeMille}}]{ShuBarGle09}
\bibinfo{author}{\bibfnamefont{E.~S.} \bibnamefont{Shuman}},
  \bibinfo{author}{\bibfnamefont{J.~F.} \bibnamefont{Barry}},
  \bibinfo{author}{\bibfnamefont{D.~R.} \bibnamefont{Glenn}}, \bibnamefont{and}
  \bibinfo{author}{\bibfnamefont{D.}~\bibnamefont{DeMille}},
  \bibinfo{journal}{Phys. Rev. Lett.} \textbf{\bibinfo{volume}{103}},
  \bibinfo{pages}{223001} (\bibinfo{year}{2009}).

\bibitem[{\citenamefont{Hunter et~al.}(2012)\citenamefont{Hunter, Peck,
  Greenspon, Alam, and DeMille}}]{HunPecGre12}
\bibinfo{author}{\bibfnamefont{L.~R.} \bibnamefont{Hunter}},
  \bibinfo{author}{\bibfnamefont{S.~K.} \bibnamefont{Peck}},
  \bibinfo{author}{\bibfnamefont{A.~S.} \bibnamefont{Greenspon}},
  \bibinfo{author}{\bibfnamefont{S.~S.} \bibnamefont{Alam}}, \bibnamefont{and}
  \bibinfo{author}{\bibfnamefont{D.}~\bibnamefont{DeMille}},
  \bibinfo{journal}{Phys. Rev. A} \textbf{\bibinfo{volume}{85}},
  \bibinfo{pages}{012511} (\bibinfo{year}{2012}).

\bibitem[{\citenamefont{Haensch}(2007)}]{T.Haensch2007}
\bibinfo{author}{\bibfnamefont{T.}~\bibnamefont{Haensch}},
  \emph{\bibinfo{title}{Metrology and fundamental constants}}
  (\bibinfo{publisher}{IOS Press}, \bibinfo{year}{2007}).

\bibitem[{\citenamefont{Ilinova
  et~al.}(2011{\natexlab{a}})\citenamefont{Ilinova, Ahmad, and
  Derevianko}}]{IliAhmDer11}
\bibinfo{author}{\bibfnamefont{E.}~\bibnamefont{Ilinova}},
  \bibinfo{author}{\bibfnamefont{M.}~\bibnamefont{Ahmad}}, \bibnamefont{and}
  \bibinfo{author}{\bibfnamefont{A.}~\bibnamefont{Derevianko}},
  \bibinfo{journal}{Phys. Rev. A} \textbf{\bibinfo{volume}{84}},
  \bibinfo{pages}{033421} (\bibinfo{year}{2011}{\natexlab{a}}).

\bibitem[{\citenamefont{Ilinova and Derevianko}(2012)}]{IliDer12b}
\bibinfo{author}{\bibfnamefont{E.}~\bibnamefont{Ilinova}} \bibnamefont{and}
  \bibinfo{author}{\bibfnamefont{A.}~\bibnamefont{Derevianko}},
  \bibinfo{journal}{http://arxiv.org/abs/1203.1963}  (\bibinfo{year}{2012}).

\bibitem[{\citenamefont{Ilinova
  et~al.}(2011{\natexlab{b}})\citenamefont{Ilinova, Ahmad, and
  Derevianko}}]{Ilinova2011}
\bibinfo{author}{\bibfnamefont{E.}~\bibnamefont{Ilinova}},
  \bibinfo{author}{\bibfnamefont{M.}~\bibnamefont{Ahmad}}, \bibnamefont{and}
  \bibinfo{author}{\bibfnamefont{A.}~\bibnamefont{Derevianko}},
  \bibinfo{journal}{Phys. Rev. A} \textbf{\bibinfo{volume}{84}},
  \bibinfo{pages}{033421} (\bibinfo{year}{2011}{\natexlab{b}}).

\bibitem[{\citenamefont{de~Araujo}(2004)}]{Ara04}
\bibinfo{author}{\bibfnamefont{L.~E.~E.} \bibnamefont{de~Araujo}},
  \bibinfo{journal}{Phys. Rev. A} \textbf{\bibinfo{volume}{69}},
  \bibinfo{pages}{013408} (\bibinfo{year}{2004}).

\end{thebibliography}

\end{document}